\documentclass[showpacs,prc,superscriptaddress]{revtex4}          
\usepackage{graphicx,epsfig}

\newcommand{ \be }{\begin{equation}}
\newcommand{ \ee }{\end{equation}}
\newcommand{ \bea }{\begin{eqnarray}}
\newcommand{ \eea }{\end{eqnarray}}

\newcommand{ \eps }{\varepsilon}
\newcommand{ \psirp }{\Psi_{RP}}
\newcommand{ \psipp }{\Psi_{PP}}
\newcommand{ \psiep }{\Psi_{EP}}
\newcommand{ \bQ }{ {\bf Q} }

\newcommand{ \epsp }{\eps_{part}}
\newcommand{ \epspp }{\eps_{PP}}
\newcommand{ \epsrp }{\eps_{RP}}

\newcommand{ \beps}{\mbox{\boldmath$\eps$}}
\newcommand{ \sgmeps }{\sigma_\eps}
\newcommand{ \sgmv }{\sigma_{vx}}
\newcommand{ \BG}{\mathrm{BG}}
\newcommand{ \G}{{\mathrm G}}
\newcommand{ \vtwo }{{v_2\{2\}}}
\newcommand{ \vfour }{{v_2\{4\}}}
\newcommand{ \bareps }{{\bar{\eps}}}
\newcommand{ \barv} {\bar{v}_2}
\newcommand{ \barx} {\bar{x}}
\newcommand{\mean}[1]{\left< {#1} \right>}

\usepackage{color}

\begin{document}

\title{
Elliptic flow in the Gaussian model of eccentricity fluctuations  
}

\author{Sergei A. Voloshin\footnote{E-mail address: voloshin@wayne.edu
(S.A. Voloshin)}}
\affiliation{Wayne State University, Detroit, Michigan 48201}
\author{Arthur M. Poskanzer}
\affiliation{Lawrence Berkeley National Laboratory, Berkeley,
California, 94720}
\author{Aihong Tang}
\affiliation{Brookhaven National Laboratory, Upton, New York, 11973}
\author{Gang Wang}
\affiliation{University of California, Los Angeles, California, 90095}

\date{\today} 

\begin{abstract}
We discuss a specific model of elliptic flow fluctuations due to Gaussian
fluctuations in the initial spatial $x$ and $y$ eccentricity components
$\left\{ \mean{(\sigma_y^2-\sigma_x^2)/(\sigma_x^2+\sigma_y^2)},
\mean{2\sigma_{xy}/(\sigma_x^2+\sigma_y^2)} \right\}$.
We find that in this model 
$\vfour$, elliptic flow determined from 4-particle cumulants,
 exactly equals the average flow
value in the reaction plane coordinate system, $\mean{v_{RP}}$, the
relation which, in an approximate form, was found earlier by Bhalerao
and Ollitrault in a more general analysis,
but under the same assumption that $v_2$ is proportional to the initial 
system eccentricity.  We further show that in
the Gaussian model all higher order cumulants are equal to $\vfour$.
Analysis of the distribution in the magnitude of the flow vector, the
$Q-$distribution, reveals that it is totally defined by two
parameters, $\vtwo$, the flow from 2-particle cumulants, 
and $\vfour$, thus providing equivalent
information compared to the method of cumulants. The flow obtained
from the $Q-$distribution is again $\vfour=\mean{v_{RP}}$.
\end{abstract}


\maketitle

\section{Introduction}

Elliptic flow is an important observable in heavy ion collision
experiments, which provides valuable information about the physics of
the system evolution starting from very early times.  Large elliptic
flow values observed recently in experiments at RHIC~\cite{Ackermann:2000tr}
are often used as an evidence for early system thermalization and as an
argument for the creation of a new form of matter, sQGP, the strongly
interacting quark-gluon plasma.  With high statistics data obtained in
the last few years at RHIC the analysis of elliptic flow becomes
dominated by systematic uncertainties, mostly by inability to separate
the so-called non-flow correlations (azimuthal correlations not
related to the orientation of the reaction plane) and the effects of
flow fluctuations~\cite{Adler:2002pu}.  Flow fluctuations can be due
to different reasons: one that has attracted much attention recently
is the fluctuations in initial eccentricity of the participant
zone. Below we discuss only the flow fluctuations related to
eccentricity fluctuations~\cite{Miller:2003kd, Manly:2005zy,
Voloshin:2006gz}.  In this paper we review the definitions of the
different coordinate systems relevant to flow analysis. Then we
discuss a particular model of eccentricity fluctuations.  Within this
model we show that by studying azimuthal correlations of produced
particles at midrapidity it is in principle impossible to separate
non-flow correlations from flow fluctuations effects as all
observables contain the same combination of the two effects.

\section{Flow coordinate systems}

\begin{figure}
\begin{minipage}[t]{0.49\textwidth}
  \includegraphics[width=.9\textwidth]{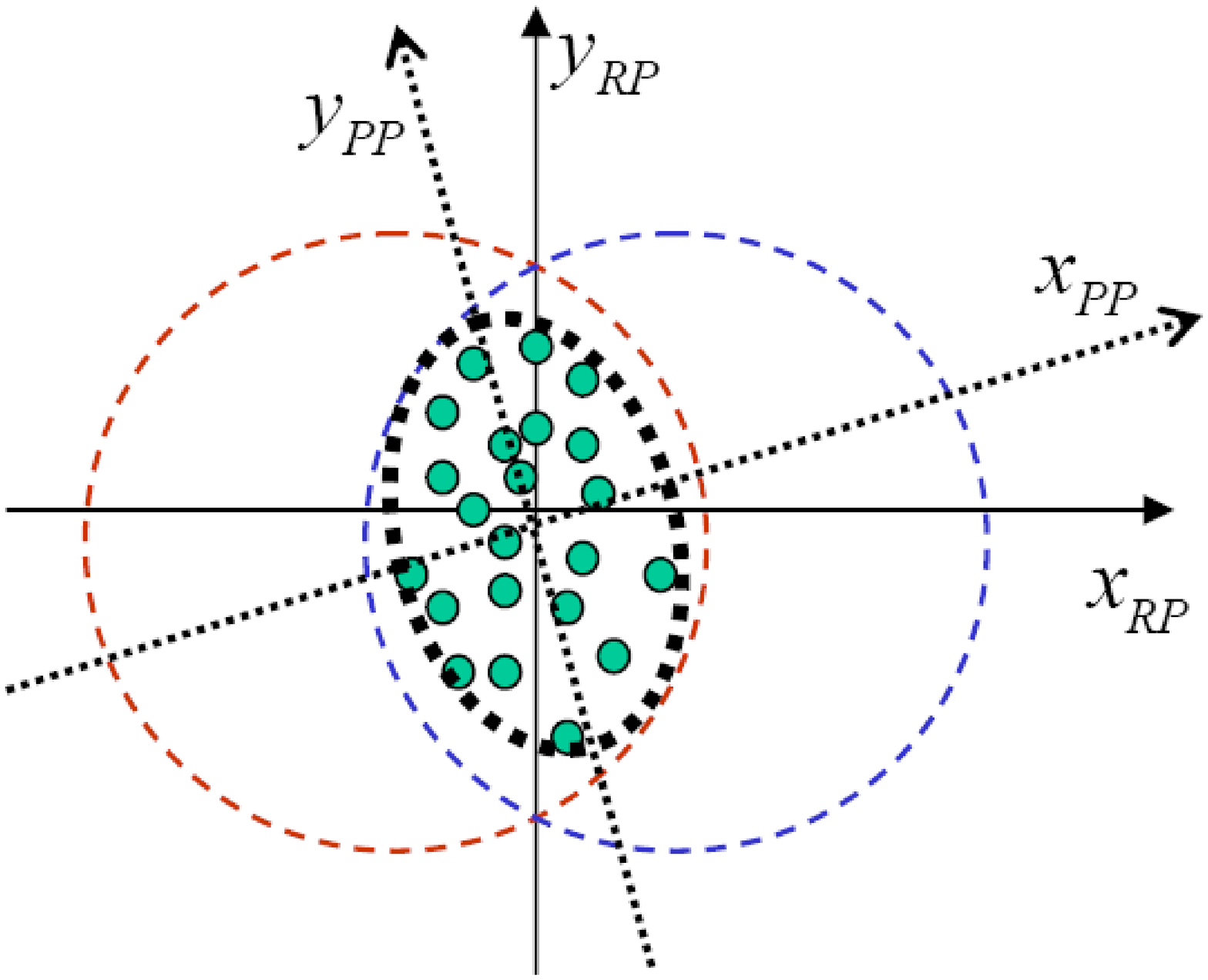} \hspace{-0.05\textwidth}
  \caption{The definitions of the $RP$ and $PP$ coordinate systems.}
  \label{coordinates}
\end{minipage}
\begin{minipage}[t]{0.49\textwidth}
  \includegraphics[width=.9\textwidth]{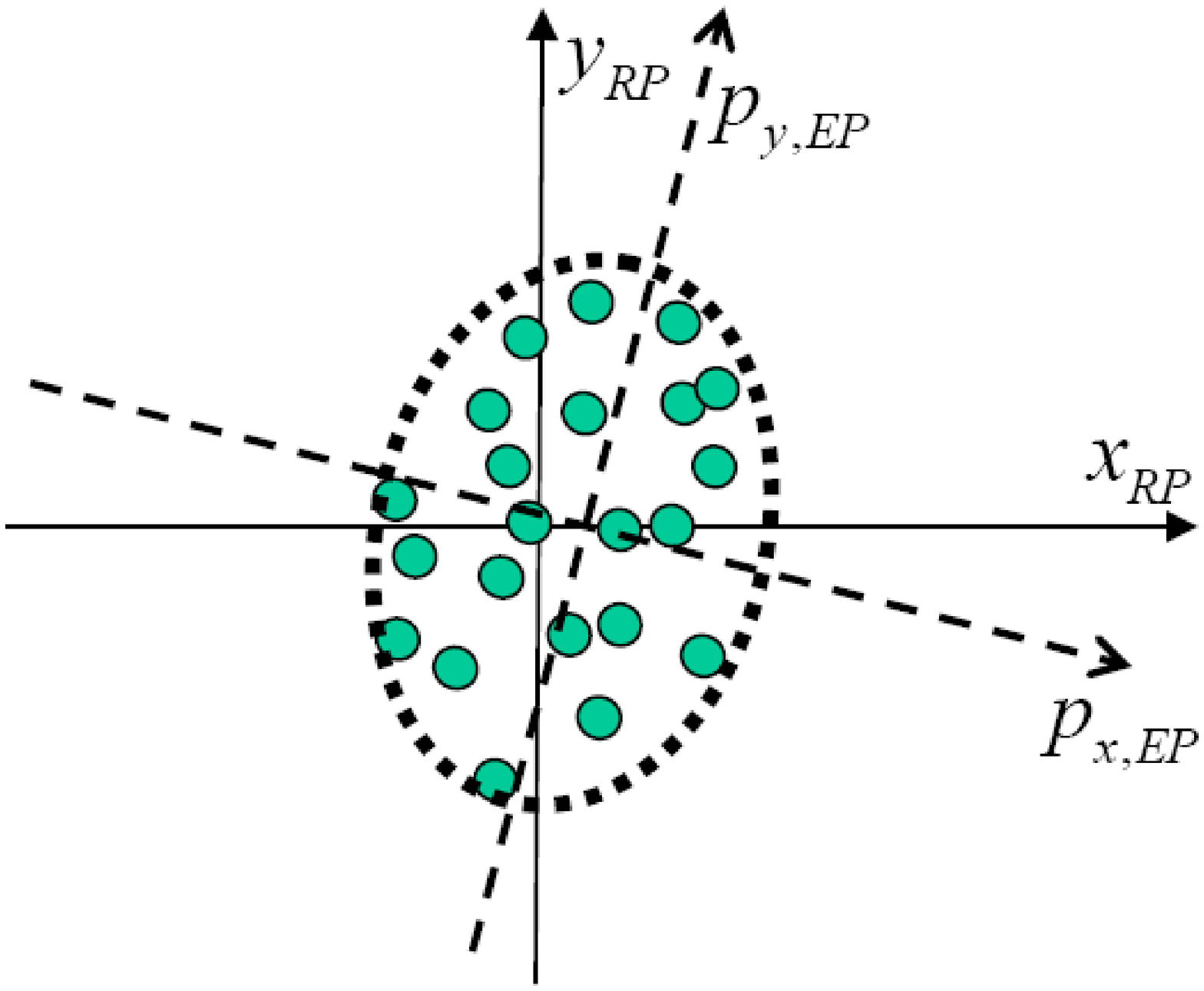}
  \caption{The definition of the $EP$ coordinate system.} 
  \label{EP}
\end{minipage}
\end{figure}

We call the coordinate system defined by the impact parameter and the
beam direction the {\em reaction plane} coordinate system, and use
subscript $RP$ to denote quantities in this system (see
Fig.~\ref{coordinates}). Then the orientation (azimuth) of the impact
parameter vector in the laboratory frame is given by $\psirp$.  The
principal axes of the participant zone will define the {\em
participant plane} coordinate system with the corresponding angle
$\psipp$, and with the $x_{PP}$ axis pointing in the direction of the
semi-minor axis of the participant zone.  We use $PP$ subscript for
quantities defined in this system.

The orientation of the flow vector $\bQ =\{Q_x,Q_y\}=\{\sum_i \cos
2\phi_i, \sum_i \sin 2 \phi_i\}$, where the sum runs over all
particles in some momentum window, defines the second harmonic {\em
event plane} (see Fig.~\ref{EP}) with corresponding azimuth $\psiep$,
$Q_x=Q\cos 2\psiep,\; Q_y=Q \sin 2\psiep$.  Although we use $Q$ in
this paper, in practice one would use $q = Q/\sqrt{N}$ in order to
minimize the effect of the multiplicity spread within a centrality
bin~\cite{Adler:2002pu}.  For a given orientation of the participant
plane, $\psipp$, anisotropic flow develops along this participant
plane.

\begin{figure}
\begin{minipage}[t]{0.49\textwidth}
  \includegraphics[width=.95\textwidth]{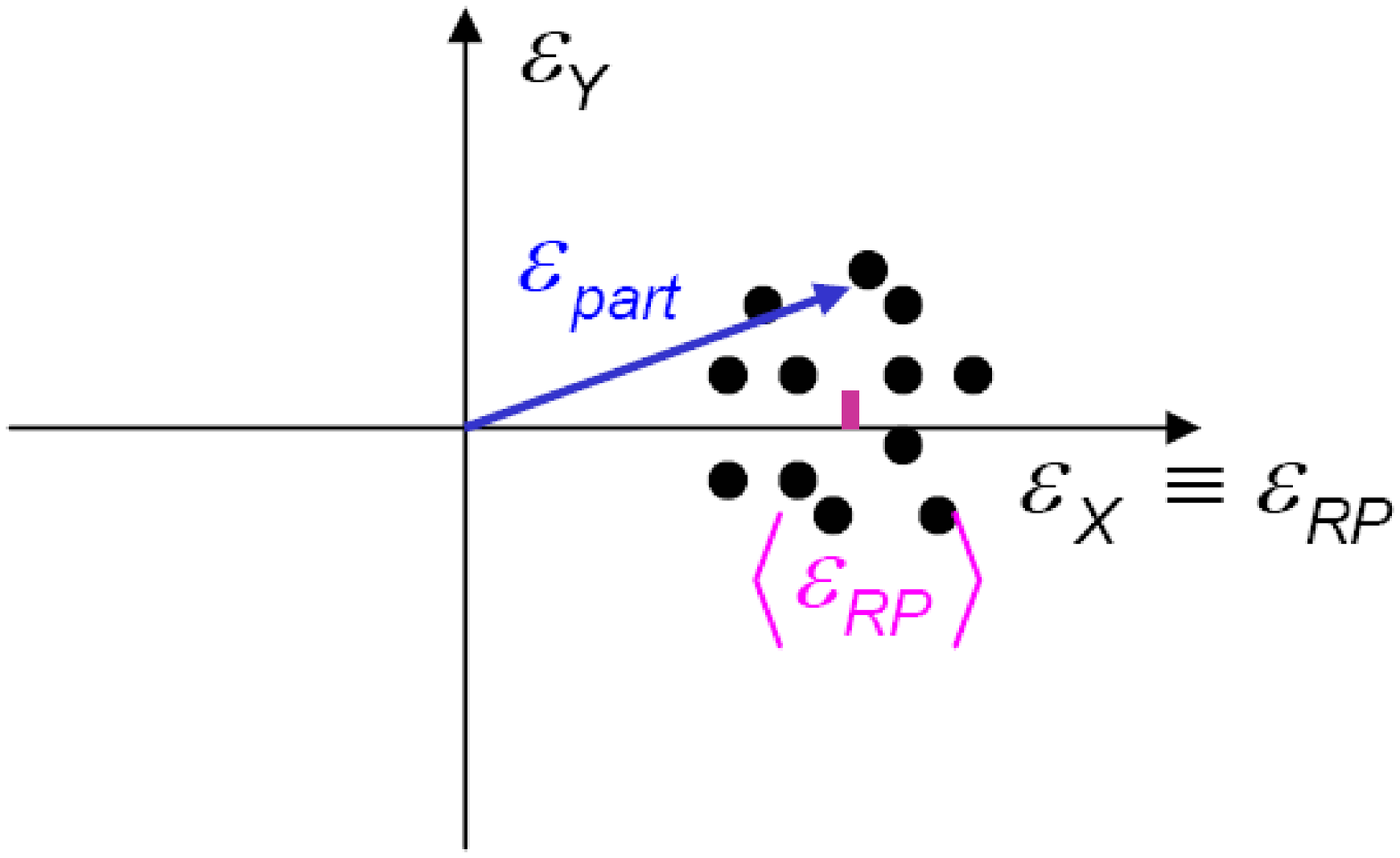} \hspace{-0.05\textwidth}
  \caption{Definition of $\epsp$.}
  \label{eps_part}
\end{minipage}
\begin{minipage}[t]{0.49\textwidth}
  \includegraphics[width=.95\textwidth]{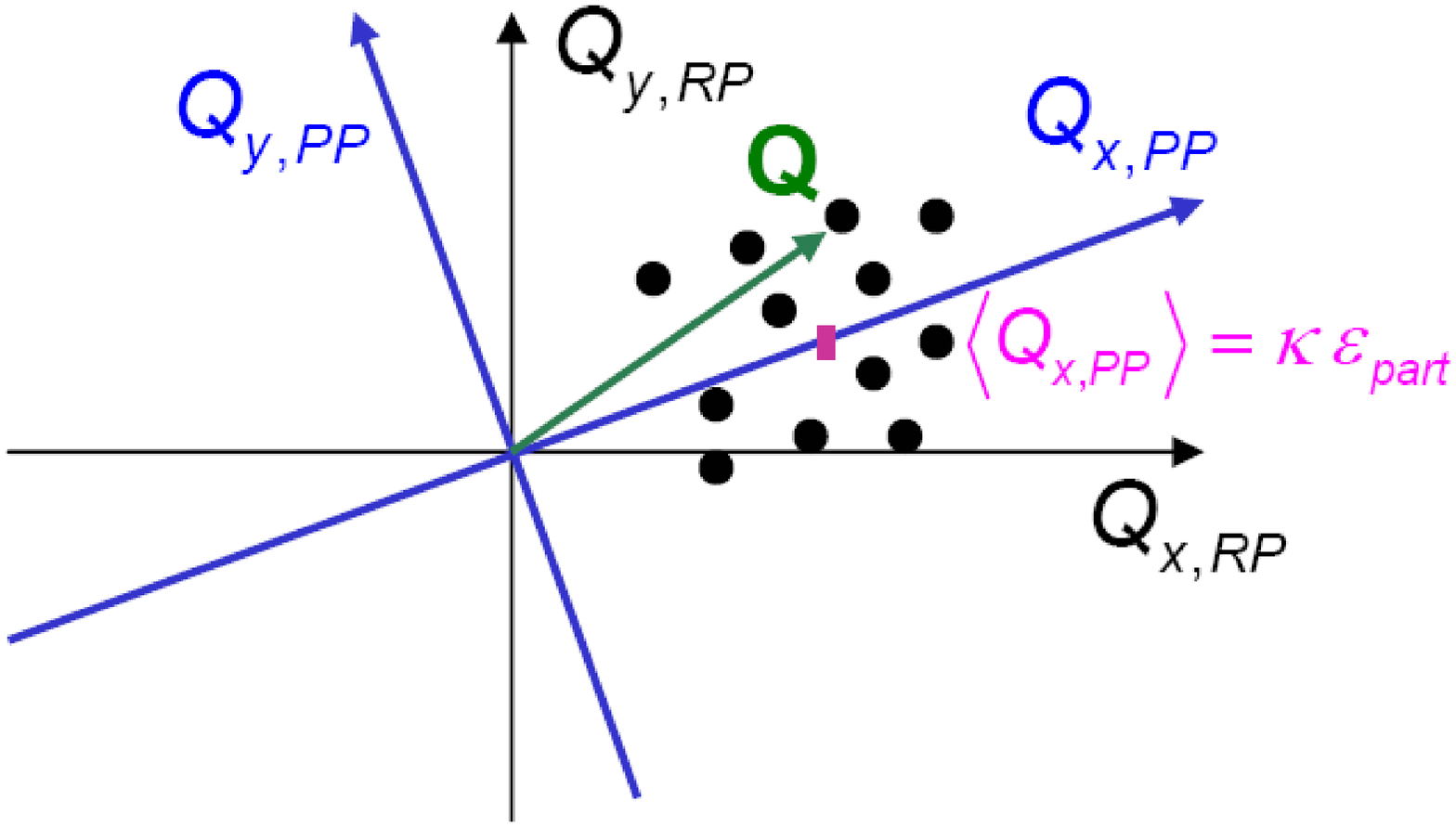}
  \caption{Flow vector distribution in events with fixed $\beps$.}
  \label{Q}
\end{minipage}
\end{figure}

The orientation of the participant plane can be also characterized by 
the {\em eccentricity vector} with coordinates
\be
  \beps=\{\eps_x,\eps_y\} =
  \left\{ \mean{\frac{\sigma_y^2-\sigma_x^2}{\sigma_x^2+\sigma_y^2}}_{part},
  \mean{\frac{2\sigma_{xy}}{\sigma_x^2+\sigma_y^2}}_{part}  \right\},
\ee
where
$\sigma_{x}^2=\mean{x^2}-\mean{x}^2$, 
$\sigma_{y}^2=\mean{y^2}-\mean{y}^2$, and
$\sigma_{xy}^2=\mean{xy}-\mean{y}\mean{x}$, 
and
 the average is taken over the coordinates of the participants in
a given event~\cite{Miller:2003kd, Manly:2005zy, Voloshin:2006gz}.
The eccentricity vector direction is given by $\psipp={\rm
atan2}(\eps_y,\eps_x)$, and its magnitude, $\epsp
=\sqrt{\eps_x^2+\eps_y^2}\equiv \epspp$, is called the {\em
participant} eccentricity (see Figs.~\ref{eps_part}, \ref{Q}) in
contrast with {\em the reaction plane (or standard)} eccentricity
$\eps_x \equiv
\epsrp$ with its mean value defined to be
\be
  \mean{\eps_x} =\mean{\epsrp} \equiv \bareps.
\ee
This mean value is approximately $\eps_{opt}$, the {\em optical}
eccentricity determined by the {\em optical} Glauber
model~\cite{Jacobs:2000wy}.

\section{Gaussian model for eccentricity fluctuations}

\begin{figure}
  \includegraphics[width=.55\textwidth]{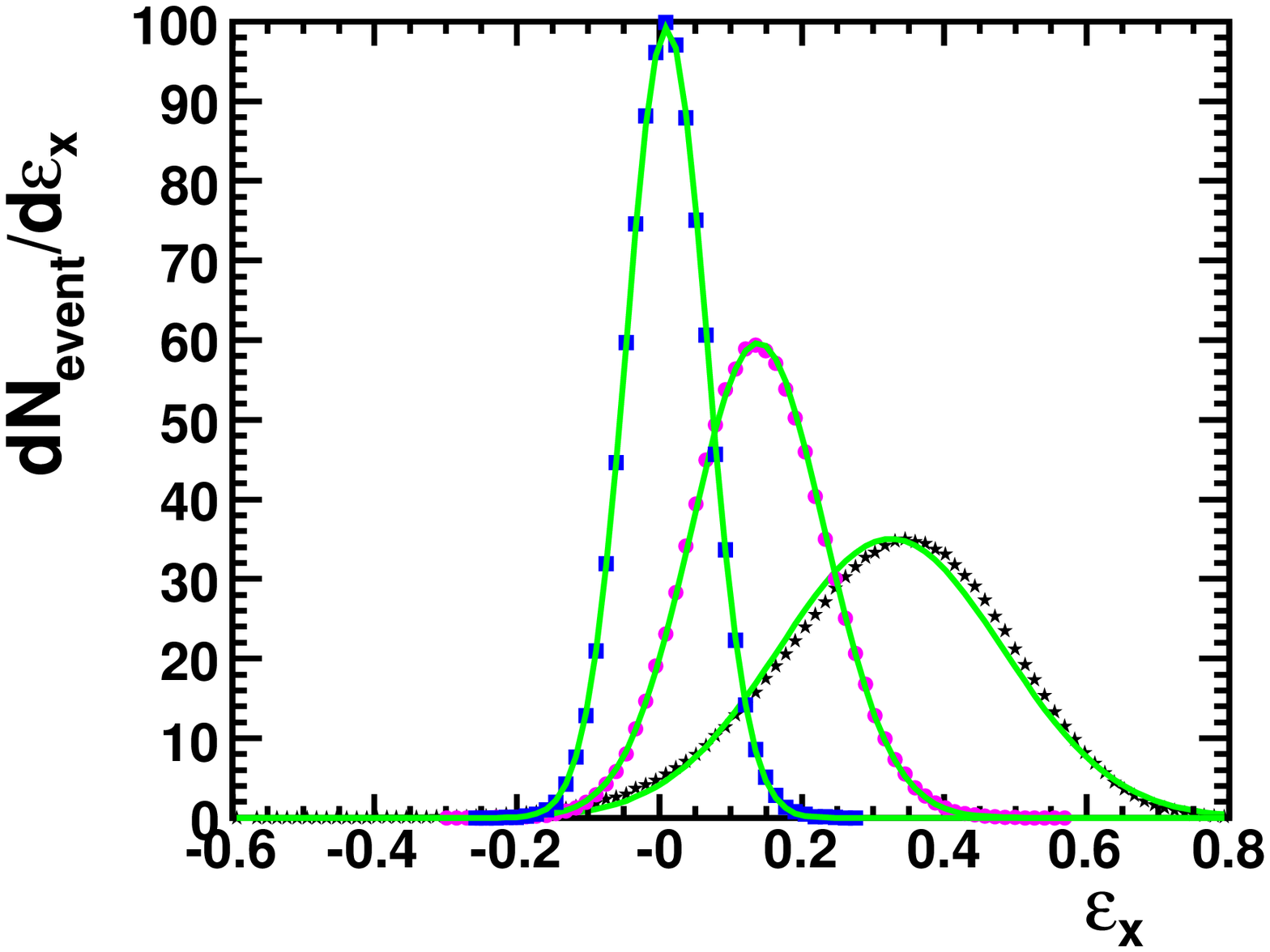}
  \includegraphics[width=.55\textwidth]{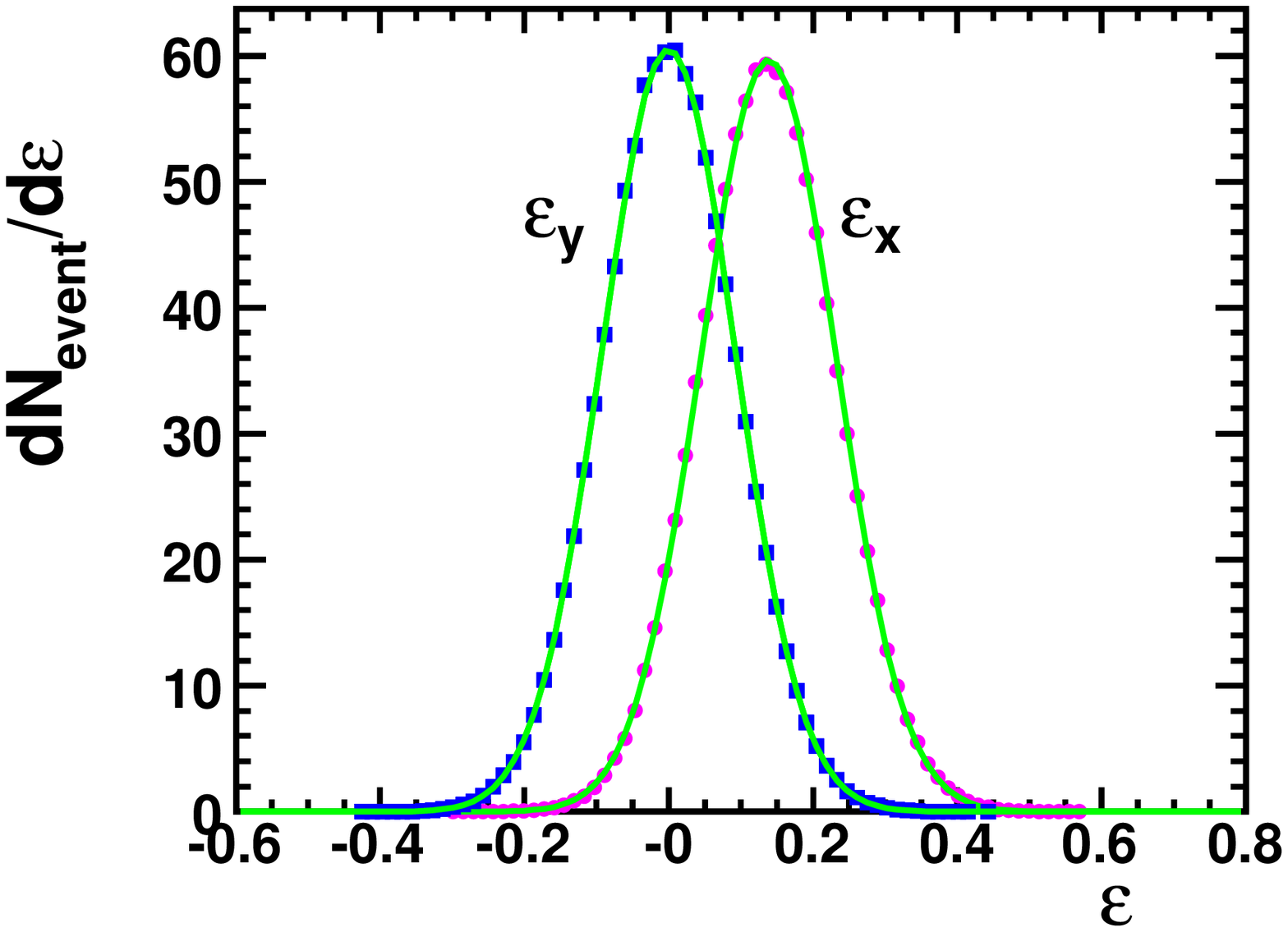}
  \caption{(top) Distribution in $\eps_x$ together with Gaussian fits
  for (left to right) central, mid-central, and peripheral
  collisions. 
	(bottom) The $y$ and
  $x$ distributions for the mid-central case. All curves have been
  normalized to the same area.}
\label{fEpsDist}
\end{figure}

\begin{figure}
  \includegraphics[width=.6\textwidth]{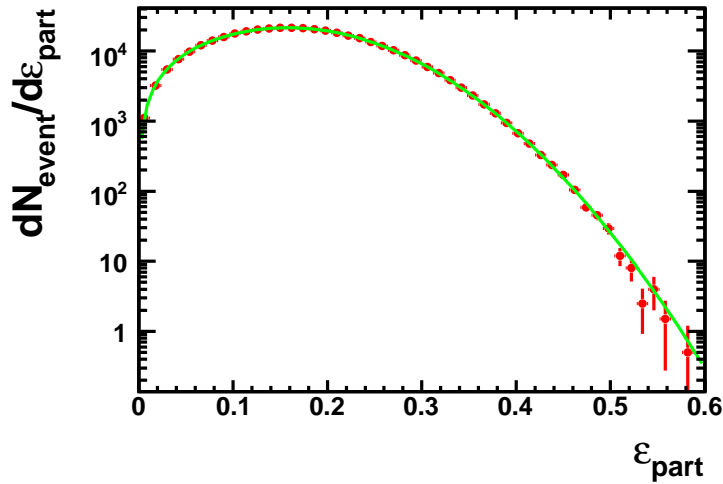}
  \caption{Distribution in $\eps_{part}$ for mid-central collisions
  ($4<b<6$~fm) and fit to the BG shape.}
\label{fEpsBG}
\end{figure}

In events with fixed $\beps$, both in magnitude and orientation, the
flow vector {\em on average} points along $\beps$, but with the
magnitude and orientation of the flow vector fluctuating due to finite
multiplicity of particles used in its definition.  As can be seen from
simulations using the MC Glauber model~\cite{Miller:2003kd,
Manly:2005zy, Voloshin:2006gz} in Fig.~\ref{fEpsDist}, the
distributions in $\eps_x$ and $\eps_y$ are well approximated by a
Gaussian form with widths approximately equal in the two directions.
There exists some deviation from a Gaussian form in peripheral
collisions, but even there the deviations are small, so we proceed
with the {\em Gaussian ansatz}.  We denote the equal widths in
$\eps_x$ and $\eps_y$ by $\sgmeps$.  The distribution in the magnitude
of the eccentricity, $\epsp$, can be obtained by integration over
angle of the vector $\beps$ as a two-dimensional Gaussian (see, for
example, the derivation in~\cite{Voloshin:1994mz}), and is given by
\be
  \frac{dn}{d\epsp}=\frac{\epsp}{\sgmeps^2} 
  I_0 \left( \frac{\epsp \mean{\epsrp} }{\sgmeps^2} \right) 
  \exp\left(-\frac{\epsp^2 +\mean{\epsrp}^2}{2\sgmeps^2}\right)
  \equiv \BG(\epsp;\mean{\epsrp},\sgmeps),
\label{eqbg}
\ee
where we have introduced a short hand notation $\BG(x;\barx,\sigma)$
for the ``Bessel-Gaussian'' distribution with one variable argument and two
constant parameters (see Fig.~\ref{fEpsBG}). Note that in $\BG(\epsp;
\mean{\epsrp},\sgmeps)$, $\epsp$ is an eccentricity as given in PP but
$\mean{\epsrp}$ and $\sgmeps$ describe the 2-D Gaussian distribution
in the RP system. The distribution is normalized to unity. For later
use we provide a few moments of the distribution
$\BG(x;\barx,\sigma)$, where $x$ is a generic variable (not the
$x$-axis):
\bea
\label{emeanx}
  \mean{x} &=& \frac{1}{2\sigma} 
  \exp\left(-\frac{\barx^2}{4\sigma^2}\right) \sqrt{\frac{\pi}{2}}
  \left[ (2\sigma^2+\barx^2)
  I_0\left(\frac{\barx^2}{4\sigma^2}\right)
  +\barx^2 I_1\left(\frac{\barx^2}{4\sigma^2}\right) \right], \\
\label{emeanx2}
  \mean{x^2} &=& \barx^2+2 \sigma^2, \\
\label{emeanx4}
  \mean{x^4} &=& \barx^4 +8 \barx^2\sigma^2 +8\sigma^4, \\
\label{emeanxsix}
  \mean{x^6} &=& \barx^6 + 18\barx^4 \sigma^2 + 72\barx^2\sigma^4+48\sigma^6. 
\eea
Note that the parameter $\sigma$ is not the variance of this
distribution; the latter would be given by
\be
\label{evariancex}
  \sigma^2_x=\mean{x^2}-\mean{x}^2,
\ee
with $ \mean{x^2}$ and $\mean{x}$ given above.  Also, from
Eqs.~(\ref{emeanx2}) and (\ref{emeanx4}) it can be shown that
\be
  2\mean{x^2}^2-\mean{x^4}=\bar{x}^4
\label{ecumx}
\ee
and
\be
  \mean{x^6}-9\mean{x^4}\mean{x^2}+12\mean{x^2}^3=4 \bar{x}^6.
\label{ecumx6}
\ee

In very central collision, the non-zero eccentricity of the overlap
region is defined mostly by fluctuations, $\mean{\epsp} \gg
\mean{\epsrp}$.  This limit corresponds to $\bar{x}
\ll \sigma$ in Eqs.~(\ref{emeanx} -- \ref{evariancex}). One finds in
this limit $\mean{x}=\sigma\sqrt{\pi/2}$ and
$\sigma_x/\mean{x}=\sqrt{4/\pi -1}$, the relation first derived
in~\cite{Broniowski}.

Figure~\ref{fEpsBG} shows the distribution in $\eps_{part}$ from the
MC Glauber calculation, together with the fit to the BG form. The
quality of the fit is good, and the extracted fit parameters shown in
Table~\ref{tbl:comp} agree well with those extracted directly from the
distributions of Fig.~\ref{fEpsDist} bottom for $\eps_x$ and $\eps_y$.

\begin{table}[hbt]
\caption{Comparison of a Gaussian distribution of $\eps$ in the RP
system with the Bessel-Gaussian fit in the PP system for
mid-central collision.}
\begin{center}
\begin{tabular}{l c c} \hline \hline
 &  $\bar{\eps}$            &  $\sigma_{\eps}$              \\ \hline   
  G, $\eps_x$, (Fig.~\ref{fEpsDist}) & 0.1384 $\pm$ 0.0001   &  0.0935 $\pm$ 0.0001 \\ 
  G, $\eps_y$, (Fig.~\ref{fEpsDist}) & 0.0000 $\pm$ 0.0001   &  0.0923 $\pm$ 0.0001 \\ 
  BG (Fig.~\ref{fEpsBG})  & 0.1344 $\pm$ 0.0002   & 0.0957 $\pm$ 0.0001 \\   
\hline \hline
\end{tabular}   
\end{center}   
\label{tbl:comp}   
\end{table}

\section{Flow fluctuations in a Gaussian model of eccentricity fluctuations}

We start our consideration by deriving the flow vector
distribution. One can approach this problem starting from two
different coordinate systems: the participant coordinate system or the
reaction plane one (see Fig.~\ref{Q}).  In the
$PP$-system the $y$ coordinate of the flow vector is not affected by
flow (and/or flow fluctuations), only the $x$ component is, which
might be taken as a simplification. On the other hand the fluctuations
in participant eccentricity (and correspondingly, in flow) have the BG
form, which is more difficult to take into account analytically.
Somewhat easier (though, obviously, equivalent) is to perform the
analysis of the $Q$-distribution in the reaction plane system.  In the
$RP$-system both components of the flow vector are affected by
eccentricity fluctuations, but the fluctuations are of Gaussian form,
with the same widths in the $x$ and $y$ directions.  Assume that
on average, flow is proportional to eccentricity with proportionality
coefficient $\kappa$:
\be
  v_2 = \kappa \epsp .
\label{ekappa}
\ee
For events with fixed $\beps=\{\eps_x,\eps_y\}$ this leads to
$\mean{Q_x}_{\beps}=N \kappa \eps_x,\; \mean{Q_y}_{\beps}=N \kappa
\eps_y$.  For the overall distribution one finds that the flow vector
is a two-dimensional Gaussian distribution with
$\mean{Q_x}=N \kappa \mean{\epsrp},\; \mean{Q_y}=0,$ and widths in
the two directions given (see~\cite{AnnRep}, \cite{Olli})
by
\bea
  \sigma^2_{Qy} = \mean{\left(\sum_i \sin 2\phi_i \right)^2}
  &=& \frac{1}{2} N [1 - \mean{\cos(4\phi_i)} + (N-1)(2 \kappa^2
  \sgmeps^2 +\delta)],
\\
 \sigma^2_{Qx} = \mean{\left(\sum_i \cos 2\phi_i \right)^2} -(N\kappa \mean{\eps_x})^2 
  &=& \frac{1}{2} N [1 +\mean{\cos(4\phi_i)} - 2 \kappa^2
  \mean{\eps_x}^2 + (N-1)(2 \kappa^2 \sgmeps^2 + \delta)],
\label{sigQxy}
\eea
where N is the number of particles, and $\delta$ is the non-flow
contribution defined by $\mean{u u^*} =
\mean{\cos(2\phi_i-2\phi_j)}=v_2^2+\delta$, 
with $u$ being the single-particle unit (second harmonic) flow vector.
Neglecting the contributions of the fourth harmonic flow and the
$(\kappa\mean{\eps_x})^2$ term, both less than or of the order of $10^{-3}$
-- $10^{-4}$ compared to unity, one finds that the widths in both
directions are the same:
\be
  \sigma^2_{Qx}=\sigma^2_{Qy} = \frac{1}{2} N [1 + (N-1)(2 \kappa^2 \sgmeps^2
  +\delta)],
\label{sigQ}
\ee
Note that $\kappa\mean{\epsrp}=\mean{v_{RP}} \equiv \bar{v}$ gives the
{\em real} flow as calculated with respect to the reaction plane and
the standard deviation of $v$ along the reaction plane axis is
$\kappa\sgmeps=\sgmv$. The distribution in flow vector magnitude would
be given then by
\be
  dn/dQ = \BG(Q;N\kappa\mean{\epsrp},\sigma_{Qx}).
\label{eQdist}
\ee  
Let us now calculate $v_2$ from 2-particle and four-particle
cumulants~\cite{Borghini:2000sa, Adler:2002pu}, $\vtwo$ and $\vfour$,
using the Gaussian ansatz for flow fluctuations.
\be
  \vtwo^2 \equiv \mean{\cos (2\phi_i -2\phi_j)}= \mean{v_2^2}+\delta=\kappa^2\mean{\epsp^2}+\delta.
\ee
Using Eq.~(\ref{emeanx2}) this becomes
\be
  \vtwo^2=\kappa^2(\mean{\epsrp}^2+2\sgmeps^2)
  + \delta=
\mean{v_{RP}}^2
+2\sgmv^2+\delta.
\label{evtwo}
\ee
Similarly, for the fourth order cumulant result, using  Eq.~(\ref{ecumx}),
\be
  \vfour^4 \equiv 2 \mean{\cos (2\phi_i -2\phi_j)}^2 - 
\mean{\cos(2\phi_i + 2\phi_j - 2\phi_k - 2\phi_m)}= 
2\mean{v_2^2}^2-\mean{v_2^4}=\barv^4=\mean{v_{RP}}^4.
\label{evfour}
\ee
Note that in this approach (Gaussian ansatz) $\vfour^4$ is always well
defined as the cumulant does not change sign.  In our model the
relation~(\ref{evfour}) is exact, but in an approximate form (and
using a different treatment of the eccentricity fluctuations) it was
derived earlier by Bhalerao and Ollitrault~\cite{Bhalerao:2006tp}, who
were the first to note that the fourth order cumulant flow
measurements are mostly unaffected not only by non-flow effects but
also by flow fluctuations.

Proceeding further, for the difference of the two cumulant results one
obtains from Eqs.~(\ref{evtwo}) and (\ref{evfour})
\be
  \vtwo^2-\vfour^2=2\kappa^2\sgmeps^2+\delta =2 \sgmv^2+\delta,
\label{ediffv2v4}
\ee
unfortunately the same parameter that defines the $Q$ distribution
width in Eq.~(\ref{sigQ}). The last observation rules out (in the
Gaussian ansatz) the possibility to measure both fluctuations and
non-flow by combining information from $Q$-distributions and
cumulants. 
Neither do higher order cumulants provide new
information. 
Using Eq.~(\ref{ecumx6}) one finds out that
\be
v_2\{6\}^6 = \left(
\mean{v_2^6}-9\mean{v_2^4}\mean{v_2^2}+12\mean{v_2^2}^3
\right) /4=
\mean{v_{RP}}^6.
\ee
One can show that in this model all higher order cumulants are given
by the corresponding power of $\mean{v_{RP}}$.  Another way to look at
this is to apply Eqs.~(\ref{ecumx}) and (\ref{ecumx6}) directly to the
$Q$ distribution Eq.~(\ref{eQdist}). One finds that the combinations
usually associated with flow cumulants~\cite{Borghini:2000sa}, are
given by corresponding powers of $Nv_{RP}$, for example
$2\mean{Q^2}-\mean{Q^4}=(Nv_{RP})^4$.

\section{Fitting $Q$-distributions}

As can be seen by comparing Eqs.~(\ref{sigQ}) and
(\ref{ediffv2v4}), $\vtwo$ and $\vfour$ completely define the
form of the $Q$-distribution, and can be used as an alternative set of
parameters compared to that in Eq.~(\ref{eQdist}). If one tries
to fit the $Q$-distribution with a functional form determined by {\em
three} parameters, e.g.  $\mean{v}$, $\sigma_v$, and $\delta$, these
parameters should satisfy the values of $\vtwo$ and $\vfour$ (which
provides only two equations), and all three can not be determined.

There can be different functional forms used to describe flow
fluctuations along the PP axis. Most often used are the Gaussian form
$\G(v;\mean{v},\sigma_v)$ and the Bessel-Gaussian $\BG(v;v_0,\sigma)$
discussed above.  Both of them have two parameters, which, as we know
can not be determined separately, so they must be correlated.

Assuming the $\BG(v;v_0,\sigma)$ form for flow fluctuations to fit the
$Q$-distribution, which would correspond to a two-dimensional Gaussian
distribution in the reaction plane coordinate system, one would find
from Eqs.~(\ref{evtwo}) and (\ref{evfour}) that the parameters are
correlated according to
\bea
  \vtwo^2 &=& const = v_0^2+2\sigma^2 + \delta \\
  \vfour &=& const = v_0.
\eea
The mean and the variance of the $v$ distribution would be given by
Eqs.~(\ref{emeanx}) and (\ref{evariancex}), but since $\sigma$ can not
be determined independent of $\delta$, $\mean{v}$ is also
undetermined.

If one uses the {\em Gaussian form} for flow fluctuations in the
$PP$-system, one would find that the parameters are correlated
according to
\bea
  \vtwo^2 &=& const=\mean{v}^2+\sigma_v^2+\delta \\
  \vfour^2 &=& const = \sqrt{\mean{v}^4-2\mean{v}^2\sigma_v^2-\sigma_v^4}
  \approx \mean{v}^2-\sigma_v^2,
\eea
or equivalently
\be
  \vtwo^2 - \vfour^2 = const \approx 2\sigma_v^2 + \delta.
\ee
The above two equations are derived in the approximation of $\sigma_v \ll
\mean{v}$ but for Gaussian fluctuations in $v$ the exact formula can
be used. Again, as $\sigma$ can not be determined independently,
$\mean{v}$ is also undetermined.

\section{Summary}

We find that in the Gaussian ansatz, fitting $Q$-distributions does
not bring any more information than that provided by cumulants.  It is
not surprising - if the distribution is defined just by two parameters
one can not get more than $\vtwo$ and $\vfour$ already provided. Note
that under this ansatz all the higher order cumulant $v$ values are
the same. The origin of the ``problem'' can be traced to the Gaussian
ansatz. It is known that for a Gaussian distribution all the cumulants
higher than rank two are zero.  The latter means that if the
collective fluctuations are of the Gaussian type one can never prove
that the fluctuations exist by any type of correlation analysis using
only particles under consideration (no external information).  A
similar problem was observed earlier in a temperature fluctuation
study of many-particle transverse momentum
correlations~\cite{multipt}. Unfortunately, deviations from a Gaussian
distribution might be too small to observe. Such deviations would show
up in the bad quality of the $Q$-distribution fits based on the
Gaussian ansatz, or in a small differences between higher order
cumulant $v$ values.

The fact that all higher order cumulants are the same and determined
by the value of flow in the reaction plane (not the participant
plane), and that that fitting of $Q$-distribution yields the same value,
explains the consistency between $\vfour$ and
$v_2\{\mathrm{ZDCSMD}\}$~\cite{GWang}, which is calculated with
ZDC-SMD as event plane and is supposed to be sensitive to $v_2$ in the
reaction plane, as well as the consistency between $\vfour$ and
$v_2\{Q-dist\}$~\cite{Adler:2002pu}.

Ref.~\cite{Wang:2006xz} used a model of flow fluctuations in which 
flow fluctuates only in the impact parameter direction.
The  use of the detectors which measure spectator neutrons, advocated
in~\cite{Wang:2006xz}, is justified for that model, but would yield zero
results for the case  of fluctuations discussed in this paper.

The authors thank P.~Sorensen for fruitful discussion. 
This work was supported in part by the HENP Divisions of the Office of
Science of the U.S. Department of Energy.

\end{document}